# Bayesian Deep Learning Approaches for Uncertainty-Aware Retinal OCT Image Segmentation for Multiple Sclerosis

Samuel T. M. Ball[1*]

*Department of Health Data Science, University of Liverpool, Waterhouse Building, Block B, Brownlow Street, Liverpool, L69 3GF, Merseyside, United Kingdom*

# Abstract

Optical Coherence Tomography (OCT) provides valuable insights in ophthalmology, cardiology, and neurology due to high-resolution, cross-sectional images of the retina. One critical task for ophthalmologists using OCT is delineation of retinal layers within scans. This process is time-consuming and prone to human bias, affecting the accuracy and reliability of diagnoses. Previous efforts to automate delineation using deep learning face challenges in uptake from clinicians and statisticians due to the absence of uncertainty estimation, leading to "confidently wrong" models via hallucinations. In this study, we address these challenges by applying Bayesian convolutional neural networks (BCNNs) to segment an openly available OCT imaging dataset containing 35 human retina OCTs split between healthy controls and patients with multiple sclerosis. Our findings demonstrate that Bayesian models can be used to provide uncertainty maps of the segmentation, which can further be used to identify highly uncertain samples that exhibit recording artefacts such as noise or miscalibration at inference time. Our method also allows for uncertainty-estimation for important secondary measurements such as layer thicknesses, that are medically relevant for patients. We show that these features come in addition to greater performance compared to similar work over all delineations; with an overall Dice score of 95.65%. Our work brings greater clinical applicability, statistical robustness, and performance to retinal OCT segmentation.

# 1. Introduction

Optical Coherence Tomography (OCT) is a non-invasive imaging technique that captures cross-sectional images of the retina[1]. Cross sectional OCT images allow for the quantitative measurement of clinically meaningful biomarkers such as the retinal nerve fibre layer (RNFL) and Ganglion cell layer and inner plexiform layer (GCL-IPL). OCT imaging is used across ophthalmology[2], cardiology[3] and neurology[4] for screening and management of health conditions. Patients with multiple sclerosis (MS) typically experience damage to the anterior visual system (e.g. the retina, optic nerves) at some point in their lives[5]. Optic nerve demyelination results in thinning of the RNFL which can be tracked via OCT imaging[6]. A related condition: optic neuritis, is also shown to be linked with MS and visible through OCT scans by highly trained ophthalmologists. This degradation of the RNFL layer over time has been shown to correlate with visual function[7] and cognitive performance[8]. Quantifying the process of RNFL thinning may help assess MS progression and inform patient care procedures for the future. Tracking of RNFL thinning can be computed from segmentation of OCT images into retinal layers, however automation of this segmentation is challenging, as OCT images have low signal to noise ratios exacerbated by recording conditions. Additionally, the subjective nature of labelling these images also presents a challenge, as differing opinions between ophthalmologists can lead to significant differences between RNFL measurements for the same patient.

The HCMS dataset[9] contains 35 individuals' right eye OCT scans; from 14 healthy controls (HC) and 21 patients with MS. Each OCT volume is split into 49 parallel consecutive B-Scans with fixed distance apart (3.9um), covering approximately 6mm x 6mm with independently reviewed manual delineations for each retinal layer (Fig. 1). Due to the open nature of the dataset, these images have become a popular benchmark for applying machine learning models (including deep learning models) to the problem of OCT segmentation[10], [11], [12], [13].

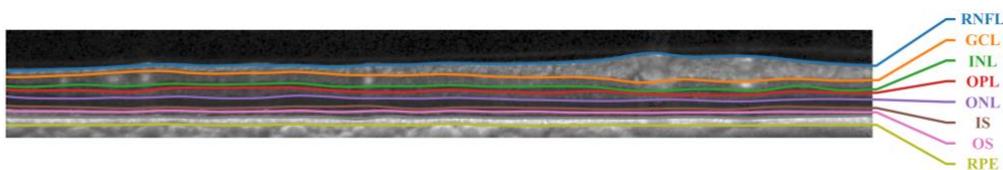

Figure 1: Example of OCT scan from HCMS dataset with labels annotated. From top to bottom: Retinal Nerve Fibre Layer (RNFL); Ganglion Cell Layer (GCL); Inner Nuclear Layer (INL); Outer Plexiform Layer (OPL); Outer Nuclear Layer (ONL); Inner Photoreceptor

Automatic OCT segmentation has seen significant recent work using neural networks, due to the flexibility of deep learning methods to adapt to image modalities. However, uptake of deep learning-based segmentation approaches experience difficulties in uptake partly due to the inability for many models to reason about the uncertainty of their estimations; process erroneous data samples outside of the training domain; or provide statistical estimates for secondary features (such as RNFL thickness) for further analysis.

In this paper, we apply the methodology of *Fully Bayesian Convolutional Neural Networks* to the HCMS dataset, allowing for uncertainty-aware segmentation of OCT imaging by modelling the weights of the network via probability distributions rather than point estimates. This allows for much richer downstream analysis compared to other methods as we see that model uncertainty can be used (for example) to identify samples with recording artefacts at inference time. Moreover, we find that our proposed method is not only more interpretable and statistically explainable than previous work but also performs better across all layers in segmentation performance.

Our contributions are therefore summarised as follows: 1) We apply a fully convolutional Bayesian neural network to an openly accessible OCT imaging dataset and show our methodology results in state-of-the-art segmentation metrics compared to previous work. 2) We show that uncertainty estimates can be used as a signal to detect erroneous images at inference time via thresholding on the total uncertainty over all pixels in a segmentation. 3) We present a novel framework for feature engineering from Bayesian segmentations, allowing for secondary features (e.g. retinal layer thickness) to have associated uncertainties.

## 2. Related Work

Deep learning methods have seen significant popularity in OCT image analysis due to the adaptive nature of neural networks to challenging modalities such as images[14]. One significant disadvantage of most neural network approaches is the inability to express uncertainty, which is critical in healthcare applications where clinical understanding of uncertainty in modelling influences patient care procedure.

Some effort has gone into modelling uncertainty in neural network outputs such as data augmentation[15] or model dropout[16] (Fig. 2); however many of these methodologies involve changes to the data or model rather than building uncertainty prediction within the model architecture itself. In other words, previous methods have resulted in *deterministic* models that are then aggregated to measure *stochastic* qualities (e.g. variance) over several data/model variations to estimate the uncertainty of the model.

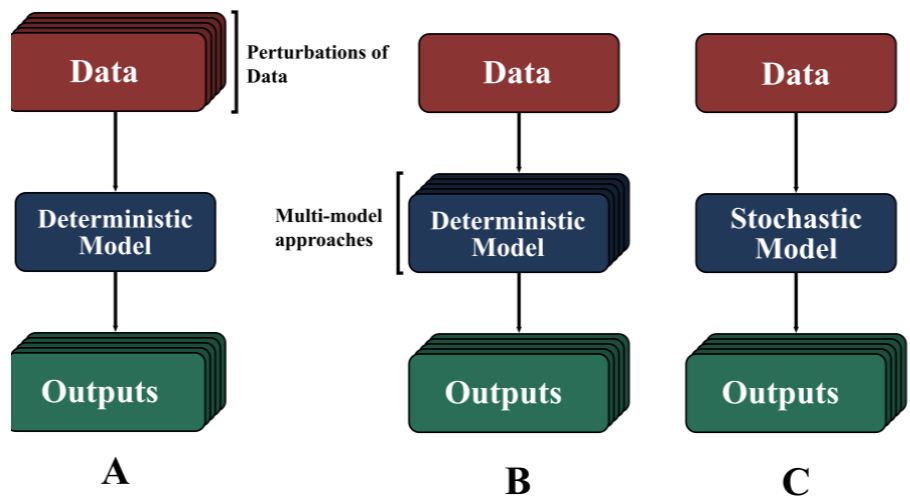

*Figure 2: Approaches to Estimating Uncertainty in Deep Learning Models. Modelling uncertainty in deep learning outputs is desirable for several reasons. Different methods of doing this include: (A) Data perturbation methods to measure how variations in data affect model outputs. (B) Multi-model approaches (e.g. ensembling, Monte-Carlo dropout) that in effect use a suite of different models and aggregate outputs for variance estimation. (C) Stochastic/Bayesian neural networks build uncertainty into the architecture itself, allowing for the model to "express" uncertainty internally rather than in post-hoc testing.*

Bayesian neural networks[17] (BNNs) are a field of research that aims to build uncertainty prediction *into* network design Bayesian neural networks function via modelling weights and biases as probability distributions rather than point estimates in training (governed by *priors* set as hyperparameters before training). This results in a *stochastic* model that can be sampled repeatedly to output a distribution of possible segmentations that can then be used to calculate uncertainty in the output given a trained model. Importantly, since these aggregated segmentations are stochastic distributions; they can be used to derive downstream distributions for measurements (e.g. RNFL thickness) for use as features in further statistical modelling in later work.

Fully Bayesian Neural networks typically[18], [19] use a normal prior for the weights in a neural network with 0 mean and diagonal covariance, for tractability and computational purposes. This specification allows the definition of loss function for training to be found as:

$$L(x|\theta) = \mathbb{E}_{\theta \sim Q}[\log P(y|x,\theta)] - \lambda D_{KL}[Q(\theta)||P(\theta)]$$

Where $y$ is an output segmentation, $x$ an input OCT image, and $\theta$ the model parameters. $\mathbb{E}_{\theta \sim Q}[\log P(Y|X,\theta)]$ is then the negative log likelihood and $D_{KL}[Q(\theta)||P(\theta)]$ the KL-Divergence between the posterior and prior model parameter distributions $Q$ and $P$. This second term acts as a regularisation parameter in training to ensure stability and avoid overfitting. $\lambda$ is a hyperparameter controlling the balance between log likelihood classification loss and KL-Divergence. Intuitively, a Bayesian Neural network is trying to minimise classification loss (as in a classical classification neural network) while preventing the model from overfitting by punishing deviation of the posterior weight distributions from the prior.

Bayesian Convolutional Neural Networks (BCNNs) extend this theory to the domain of image processing, using probabilistic convolutional filters to derive feature maps rather than traditional point estimate kernels. Such BCNNs have previously been applied to segmentation and anomaly detection for OCT imaging[20], [21], [22], [23], [24], allowing for visualisation of uncertainties around OCT image segmentation – however the specific implementation of the Bayesian component is typically achieve by some form of Monte Carlo dropout. In traditional Bayesian methods, these uncertainties can be split into *epistemic* and *aleatoric* uncertainties: Epistemic uncertainties represent uncertainties resulting from the model, whereas aleatoric uncertainty captures inherent randomness within the data. It can be shown[25], [26] that the uncertainties in fully Bayesian Neural Networks can be split via the formulae:

$$\sigma^2_{ale}(p(y|x)) = \frac{1}{T}\sum_{t=0}^{T} \text{diag}(\widehat{p_t}) - \widehat{p_t}\widehat{p_t}^T \qquad (1)$$

$$\sigma^2_{epi}(p(y|x)) = \frac{1}{T}\sum_{t=0}^{T} (\widehat{p_t} - \bar{p})(\widehat{p_t} - \bar{p})^T \qquad (2)$$

$$\sigma^2_{total}(p(y|x)) = \sigma^2_{epi} + \sigma^2_{ale} \qquad (3)$$

Where $\widehat{p_t} = \text{Softmax}(f_{\theta_t}(x))$, $f$ being the neural network, $x$ the inputs, $y$ the output segmentation, and $\bar{p} = \frac{1}{T}\sum_{t=0}^{T}\widehat{p_t}$, where $t$ represent each sample from the Bayesian neural network.

The splitting of epistemic and aleatoric uncertainties for analysing biomedical imaging can be used to further reason about models in practice. Previous work (using Monte Carlo Dropout methods) has shown the usage of epistemic uncertainty as a method of detecting anomalous samples for detection of age related macular degeneration[24], however to our knowledge no work has used Fully Bayesian networks directly to achieve this.

# 3. Methods

## 3.1. Data Preparation, Preprocessing and Mask Generation

The 14 healthy controls and 21 MS patients (each containing 49 B-Scans) were randomly each split into 80% training and 20% testing sets, then combined to ensure even distribution of healthy controls and MS patients in both the training and validation sets. For segmentation, masks were generated from the delineations for each two-dimensional OCT image slice. These images and masks were then resized to 512x512 pixel images for use for input into the deep learning model (Fig. 3). Inputs were normalised for a mean pixel value of 0.5 and variance 0.5 to ensure consistent brightness across all samples.

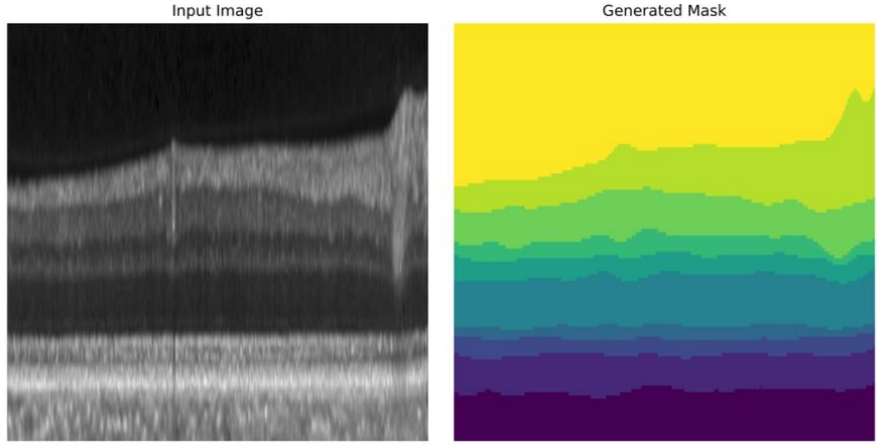

*Figure 3: Example of image with corresponding mask used for input and comparison for neural network training. Masks were generated from the manual delineations by taking areas between pairs of labelled lines (e.g. RFNL-GCL). Images and masks were subsequently downscaled to 512x512 for input into the neural network model.*

## 3.2. Bayesian Neural Network

In this work, a BCNN model based on the U-Net architecture[27] was created for segmentation of the OCT images. This model is similar to the original U-Net model with all convolutional units replaced with Bayesian Convolutional layers, and all transposed convolutional (i.e. deconvolutional) layers replaced with Bayesian implementations. The hyperparameters used for training can be seen in Table 1. Training was implemented using Pytorch[28] using a NVIDIA RTX 4000 Ada GPU with 20GB memory. During training, the ELBO loss for both training and validation sets, and Dice scores for the validation dataset were recorded.

*Table 1: Hyperparameters for model training.*

| Hyperparameter | Value |
| --- | --- |
| Image Size | 512x512 |
| Batch Size | 4 |
| Epochs | 10 |
| Learning Rate | 1e-4 |
| KL Divergence Weight ($\lambda$) | 1 |

### 3.3. Model Validation

After the model was trained, for comparison to similar work the model then sampled 64 segmentations for each of the test images and then averaged probabilities to give mean pixel-wise classification for each image – the maximal mean probability for each pixel was taken as the corresponding mask class label for the overall model for point estimate segmentation:

$$\overline{y_{i,j}} = \underset{c}{\mathrm{argmax}} \left( \frac{1}{T} \sum_{t=0}^{T} p_{t,c,i,j} \right)$$

Where $i, j$ represent spacial dimensions of the image; $c$ the different mask classes (e.g. background, RNFL) and $t$ the different samples from the neural network. These classifications were then compared to the ground truth masks from the original dataset using the Dice score after resizing the predictions back to their original size (1024 by 128px):

$$Dice(y_{true}, y_{pred}) = \frac{2|y_{true} \cap y_{pred}|}{|y_{true}| + |y_{pred}|}$$

Some previous work aggregates some layers together (e.g. GCP-IPL, INL-OPL and ONL-IS-OS) due to differences in opinion as to how OCT scans should be analysed. For comprehensive comparison, we also record Dice scores for the mapping of these classes together (i.e. – relabelling all IPL labels in both ground truth and predictions as GCP).

### 3.4. Uncertainty Analysis

After validating the model using the Dice scores, for each image (training and validation) we sample the model 64 times to get a distribution of pixel-wise segmentations for each class. We calculate the epistemic and aleatoric uncertainty (via variance) for each image using (1-3), as well as the total uncertainty across all pixels in an image for each class.

Using the point estimates in conjunction with the total uncertainty components, we analyse both *areas* of images with high uncertainty and images where *total uncertainty* is highest. The former can be used to qualitatively identify potentially problematic sections of an image where the segmentation may be inaccurate that may require additional input from clinical experts; whereas the latter can be used to identify *entire images* that exhibit high uncertainty where the image may need to be retaken.

*Table 2: Dice scores for our Bayesian model versus other deep learning work on the OCT-MS Dataset. We find our method performs similarly to other methods in the field, improving in accuracy in some areas and less in others. A single asterix(*) notes that the metric includes the next layer(s) in its calculation in the original work. We also note that due to different mask generation methods, the RPE measurements are significantly different. Additionally, we record our method with aggregated layers (as in some other work) for more accurate, but less granular classifications.*

| Layer | RNFL | GCP | IPL | INL | OPL | ONL | IS | OS | RPE | Average |
|---|---|---|---|---|---|---|---|---|---|---|
| Wang et al[29] | **94.1** | 95.4* | - | 88.5 | 89.4 | 94.9 | 88.2 | 89.0 | 92.3 | 90.9 |
| Lopez-Varela et al[30] | 88.8 | 75.7 | 81.1 | 83.5 | **91.1** | 88.4 | 83.5* | - | 94.0 | 86.0 |
| Chen et al[31] | 92.4±2.1 | 94.06±1.8* | - | 93.9±1.4* | - | 96.0±1.6* | - | - | 93.9±0.9 | 94.0±1.5 |
| Chang et al[32] | 93.7 | **95.5*** | - | 89.1 | 91.0 | 95.7 | 89.8 | 91.0 | 94.5 | 92.1 |
| He et al[33] | 93.7 | 95.1* | - | 88.3 | 90.2 | 95.0 | 87.2 | 87.2 | 90.7 | 90.33 |
| Proposed | 93.1±4.0 | 89.7±4.1 | **88.4±3.6** | 95.2±1.5 | 90.8±2.8 | 87.9±3.7 | **95.4±2.1** | **94.4±2.4** | 99.2±0.4 | 92.7±2.5 |
| Proposed (agg) | 93.1±4.0 | 93.1±4.0 | - | **96.7±1.0** | - | **98.0±0.1** | - | - | **99.2±0.4** | **95.65±1.5** |

# 4. Results

After training the Bayesian segmentation model for 100 epochs, we graph the loss over each epoch in Fig. 4. We see that the loss continues to decrease in both training and validation datasets; however, the validation Dice score stays constant throughout training (from 0.925 to 0.945). Dice score comparisons of the final model versus similar work can be seen in Table 2. Fig. 5 shows qualitative examples of original images, ground truth masks and predictions.

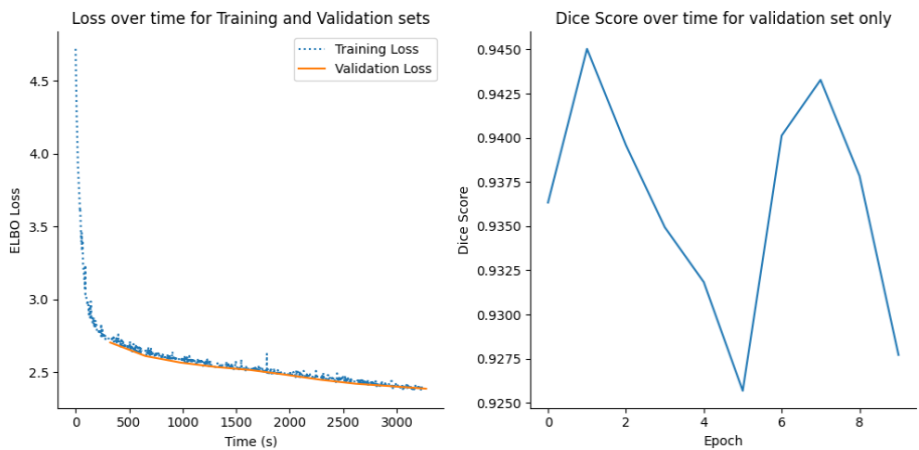

*Figure 4: Training and Validation Loss; and Validation Dice scores for training Bayesian U-Net Model on HCMS OCT dataset. We find that the model quickly converges to achieve high Dice scores (right) on the validation set after the first epoch – however the loss still decreases throughout training. The small variation in the Dice Score may suggest this decreasing loss comes from decreasing KL divergence (i.e. regularisation) rather than improvements in classification.*

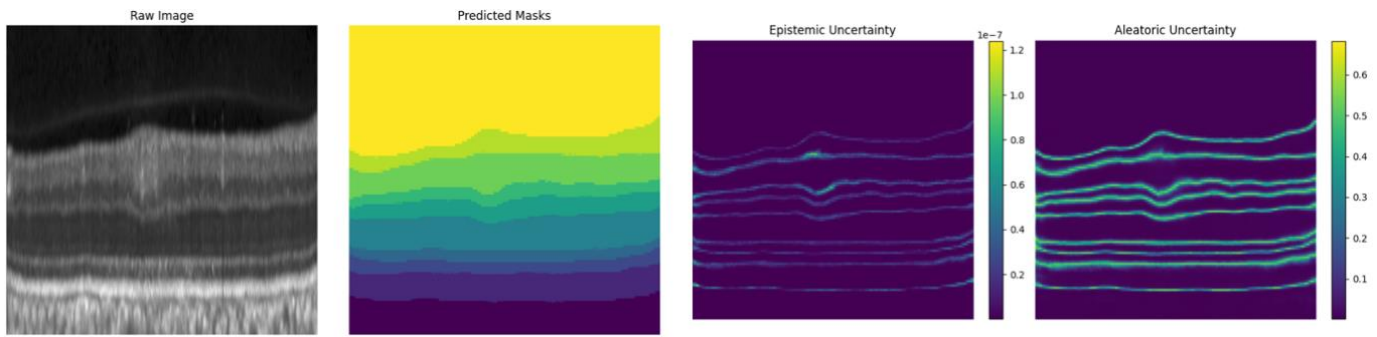

*Figure 5: Uncertainty Estimates for an example output and image from the Bayesian model. We see that most of the uncertainties are aleatoric (note the difference in scale bars); and most of this uncertainty occurs at the boundaries of class labels. This may be to be expected, particularly in a modality such as OCT imaging where signal to noise ratio is especially low; with boundary delineations between retinal layers particularly "fuzzy".*

## 4.1. Uncertainty Analysis

A unique property of our proposed method is the ability to reason about uncertainty using Bayesian Neural Networks in a way that traditional neural networks are unable to do. Fig. 6 qualitatively shows a sample image output mask along with epistemic and aleatoric uncertainty values. The relatively high aleatoric uncertainty (>99.99% proportion) suggests that the model is highly confident about its parameters; with the remaining uncertainty comes from variation within the data itself.

This feature of the model is especially helpful in identifying samples that may exhibit anomalous features and artefacts without relying on manual annotation. Fig. 7 shows the total uncertainty across all masks, per slice, per sample. We see that by setting a total uncertainty threshold of 23000, we identify 3 samples with significant errors in recording that would be candidates for reimaging later. These samples include misaligned images, images with shadows and images with significant noisy elements throughout.

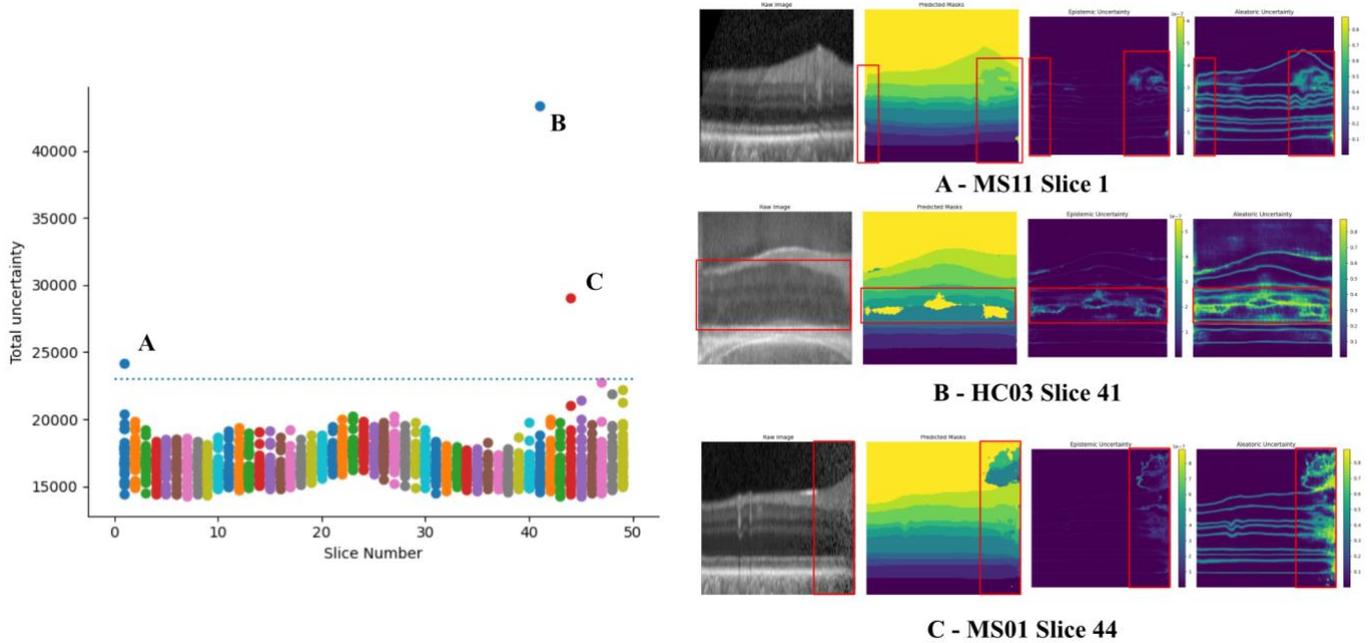

Figure 6: Uncertainty estimates of each slice for each patient. By thresholding the total uncertainty in an image (left, dotted line at 23,000), we can identify images with significant artefacts or errors in processing at inference time, independent of ground truth labels. In the first slice for the MS11 sample (A), we see significant uncertainty due to OCT recording noise in the right and left of the image (circled in red). In the 41st slice for the HC03 sample, we see that reduced contrast in the OCT scan (likely due to miscalibration) results in the inability to delineate the layers easily. We see that this creates some confusion for the model, with significant uncertainty in the middle third of the image (circled in red). Finally, in the 44th slice for sample MS01, we see increased noise (likely due to misalignment) of the OCT image which causes some uncertainty on the right-hand side of the image (highlighted).

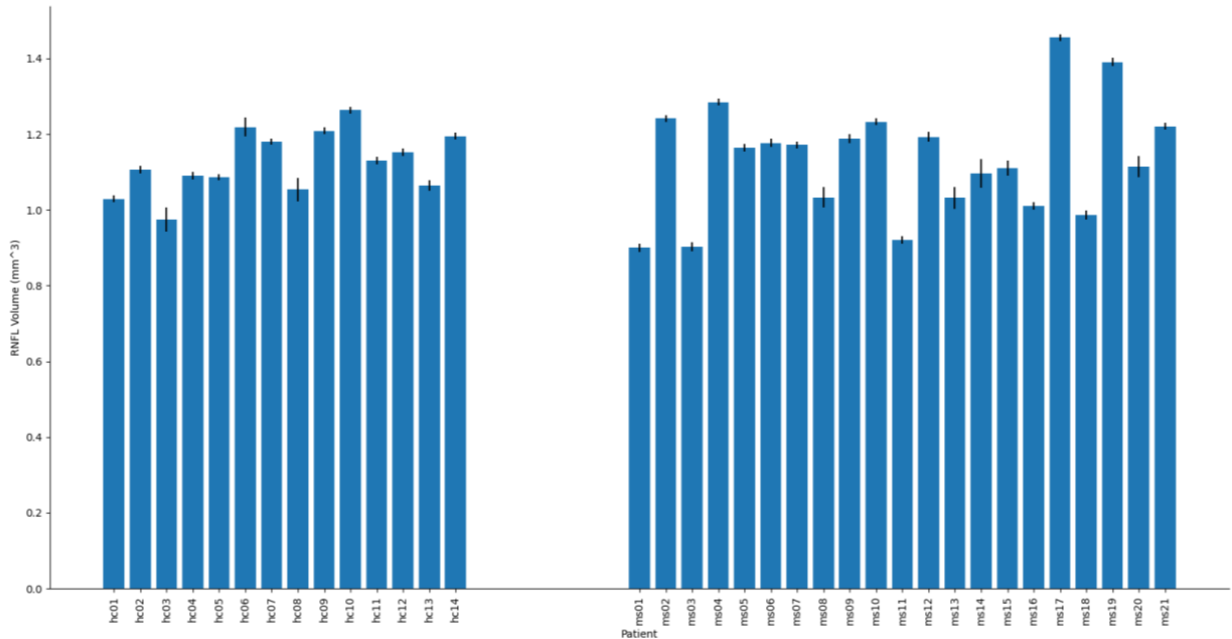

*Figure 7: Total RNFL Mask Areas (i.e. Volumes) for each patient. We see that significant variance within the two groups (HC and MS) means it is challenging to see the effect of MS on RNFL thickness within OCT imaging. This may be due to slight variations in the angle or focus of the OCT scans; or artefacts within the OCT scans, or perhaps the effect of MS on RNFL thickness is more spatially complex (e.g. changes in the shape, rather than volume) that needs further investigation.*

# 5. Discussion

### 5.1. Dataset Size and Computational Cost

The relatively low n=35 patients in the dataset is an interesting point for deep learning specialists – common wisdom in the field is that large sample sizes are required to train a model accurately. In some ways, one of the main challenges of the dataset analysed in this work is achieving high accuracy segmentations *without* large volumes of high-quality data. In this work (as with the other work compared) we have showed state-of-the-art results with a low sample size are possible with intelligent model design on consumer hardware, despite the small sample size. Crucially, however this performance can come *in addition* (rather than at the cost of) to the ability to reason statistically about outputs via sampling the model repeatedly, a feature highly desirable in medical imaging fields. A larger sample size for this would also allow for a more accurate classification model to be trained (e.g. a Bayesian ResNet model) – as currently the low sample size makes this challenging.

### 5.2. RNFL Thickness Measurement as an biomarker for MS

RNFL thickness has been used as a biomarker to assess retinal degradation and cognitive function via traditional machine learning and deep learning methods[34], [35], [36]. With Bayesian neural networks; we can derive the RNFL thickness by taking point estimates of the RNFL mask total areas (combining to make a volume across the retina) and totalling the uncertainties over the masks for an uncertainty estimate. Unfortunately, we find (Fig. 7) that there is little association between the derived RNFL thicknesses from the model and MS diagnosis alone, but a relatively high uncertainty when calculating these volumes.

We hypothesise that OCT imaging's low signal to noise ratio means it may not be a sensitive enough modality to pick up changes in RNFL morphology for MS patients by itself. Variance within the OCT scans (e.g. angle, focus, existence of artefacts such as shadows) may contribute more to RNFL thickness as a biomarker via OCT imaging than any effect of MS itself. Further work with a larger sample of MS patients may be able to discriminate between MS and healthy control patients using RNFL thickness.
Furthermore, some previous work[36] shows using the RNFL thickness alone might not be sufficient to screen for MS, as it correlates more strongly with cases of optic neuritis. We hope that this work may be used as a richer biomarker and allow for better quality control for these studies.

### 5.3. Choice of $\lambda$

$\lambda$; the regularisation loss associated with ensuring the probability distributions are close to the priors, is thought to counteract overfitting at the cost of performance. However, we find little effect on validation performance based on the choice of $\lambda$ (Fig. 8), with the training loss decreasing faster, later at higher values of $\lambda$, further suggesting that this late-stage improvement is due to the regularisation loss rather than improvements in pixel classification. Indeed, when $\lambda = 0$, we see no further improvement in the combined loss at the end of training, suggesting pixel classification convergence.

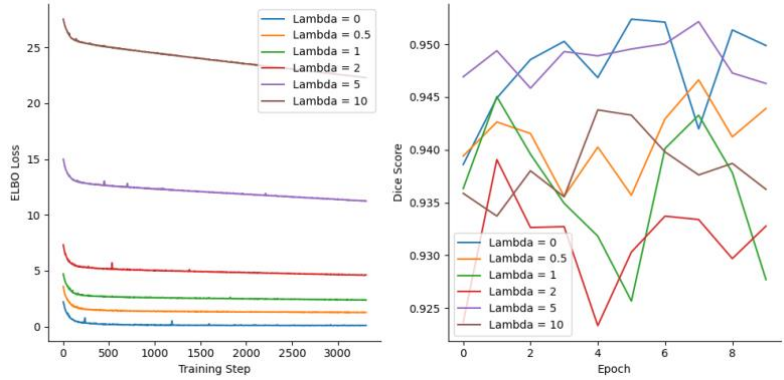

*Figure 8: Training Losses and Validation Dice Scores for Different Regularisation constant values of λ: We see that as the value of λ increases, the gradient of the training progress after the initial epoch also increases (left) despite no noticeable change in the validation Dice score (right). This suggests that after the first epoch, the model is mainly regularising the model rather than learning new features.*

## 6. Conclusion

In conclusion, this study demonstrates the significant advantages of applying Bayesian convolutional neural networks to the delineation of retinal layers in OCT scans. Our findings reveal that Bayesian models not only outperform traditional non-Bayesian methods in terms of accuracy across all layers averaged, achieving a state-of-the-art Dice score of 95.65%, but also provide crucial uncertainty estimates that enhance the reliability and interpretability of the results. The ability to identify erroneous samples and derive biomarkers with associated uncertainties represents a substantial improvement over existing approaches, addressing the statistical maturity concerns that have hindered the widespread adoption of automated OCT analysis, although more work is needed with a larger sample to utilise these biomarkers.

The integration of uncertainty reasoning in our Bayesian model offers a more robust framework for clinical decision-making, potentially reducing the time and bias associated with manual delineation. This advancement has broad implications for the fields of ophthalmology, cardiology, and neurology, where OCT is extensively used. Future work should focus on validating these findings across larger and more diverse datasets, as well as exploring the clinical integration of these models to assess their real-world impact.

Overall, our work underscores the potential of Bayesian deep learning models to revolutionize the automation of OCT image analysis, providing more reliable and efficient diagnostic tools that can significantly benefit both clinicians and patients.